\documentclass[aps,pra,preprint,showpacs]{revtex4-1}
\usepackage{amsmath,graphicx,epsfig}
\begin{document}
\def\la{{\langle}}
\def\ra{{\rangle}}
\def\q{{\quad}}

\title{Are the 'weak measurements' really measurements? }
%
%
\author {D. Sokolovski}
\affiliation{Departamento de Qu\'imica-F\'isica, Universidad del Pa\' is Vasco, UPV/EHU, Leioa, Spain}
\affiliation{IKERBASQUE, Basque Foundation for Science, E-48011 Bilbao, Spain}

   \date{\today}
   \begin{abstract}
'Weak measurements' can be seen as an attempt at answering the 
'Which way?' question without destroying interference between the pathways involved.
Unusual mean values obtained in such measurements represent the response of a 
quantum system to this 'forbidden' question, in which the 'true' composition of virtual pathways
is hidden from the observer. Such values indicate a failure of a measurement
 where the uncertainty principle says it must fail, rather than provide an additional insight into physical reality.

\end{abstract}

%
%
\maketitle

{\it Q. What time is it when the clock strikes 13?}

{\it A. Time to buy a new clock.}

(A joke)

\section{Introduction}
Twenty five years ago Aharonov, Albert and Vaidman published a
paper entitled "How the Result of a Measurement of a Component of the Spin of a Spin-1/ 2 Particle Can Turn Out to be 100?"  \cite{W1}.
The authors' idea was further developed a large volume of work on the so-called 'weak measurements' (see, for example, \cite{W2}-\cite{W10}), culminating in a somewhat bizarre the BBC report \cite{BBC} suggesting that 
 "Pioneering experiments have cast doubt on a founding idea of the branch of physics called quantum mechanics".
 There seems to be room for discussion about what actually happens in a 'weak measurement', and this is the subject of this paper. Some of the early and more recent criticisms of the original approach used in Ref. \cite{W1}  can be found in Refs. \cite{C1}-\cite{C5}.
 \newline
There appear to be only two possible answers to the original question posed by the authors of Ref.\cite{W1}:
(I) there is a new counter-intuitive aspect to quantum measurement theory, or (II) the proposed measurement is flawed.
In this paper we will follow Ref. \cite{C3} in advocating the second point of view.
The argument is subtle. There is no error in the simple mathematics of the Ref. \cite{W1}.
It is the interpretation of the result which is at stake. 
Below we will argue that a  'weak measurement' attempts to answer the 'Which way?' question without destroying
interference between the pathways of interest. Such an attempt must be defeated by the Uncertainty Principle 
 \cite{Feyn}, \cite{Bohm} and the unusual 'weak values' are just the evidence of the defeat.

\section {Probabilities and  'negative probabilities'}
A random variable $f$ is fully described by its probability distribution $\rho(f)$.
Often it is sufficient to know only the typical value of $f$, and the range over which 
the values are likely to be spread.
To get an estimate for the centre and the width of the range, one usually evaluates the mean value of $f$,
$$\la f\ra=\int f \rho(f)df /\int \rho(f)df$$ 
and the standard mean deviation 
$\sigma=\sqrt{\la f^2\ra-\la f\ra^2}$. 
\newline
 Suppose $f$ can only take the values $1$ and $2$, and 
its unnormalised probability distribution is  $\rho(1)=1.1$ and $\rho(2)=1$. 
We, therefore, have
\begin{eqnarray} \label{N1}
\la f\ra=[1\times \rho(1)+ 2\times \rho(2)]/[\rho(1)+\rho(2)]\approx 1.4761,\\ \nonumber
\sigma
\approx  0.4994,
\end{eqnarray}
which reasonably well represent the centre and width of the interval $[1,2]$
containing the values of $f$. 
\newline
Suppose next that, for whatever reason, the unnormalised probabilities were
allowed to take negative values, e.g.
\begin{equation} \label{N2}
\rho(1)=-1.1, \q \rho(2)=1.
\end{equation}
Using the same formulas, we find
\begin{eqnarray} \label{N3}
\la f\ra= -9, \q\q \sigma \approx 10.49i.
\end{eqnarray}
which, clearly, no longer describe the range $[1,2]$, - $|\la f\ra|$ is too large, 
and $\sigma$ is purely imaginary. The reason for obtaining such an 'anomalous' mean value is that the denominator 
in Eq.(\ref{N1}) is small, while the numerator is not - hence the large negative 'expectation value' 
in Eq.(\ref{N3}).
\newline
In general, the mean and the standard mean deviation of an alternating 
distribution do not have to represent the region of its support.
These useful properties of $\la f\ra$ and $\sigma$ are lost, once a 
distribution is allowed to change its sign.

\section {Complex valued distributions}

To make things worse, let us assume that the unnormalised
'probabilities' $\rho(f)$ are also allowed to take complex values,
\begin{equation} \label{C1}
\rho(f)=\rho_1(f)+i\rho_2(f),
\end{equation}
while $f$ may take any value inside an interval $[a,b]$.
As before, we will construct a normalised distribution
$w(f)\equiv \rho(f)/\int_a^b \rho(f')df'$, which can now be written as a sum
 of its real and imaginary parts,
\begin{eqnarray} \label{C2}
w(f)\equiv w_1(f)+iw_2(f) = \quad \quad \quad \quad \quad \quad \quad \quad \quad \quad \quad \quad \quad \quad \quad \quad \\
\nonumber
\\ \nonumber
 \frac{A_1^2 (\rho_1(f)/A_1)+A_2^2 (\rho_2(f)/A_2)}
{A_1^2+A_2^2}+iA_1A_2\frac{\rho_2(f)/A_2-\rho_1(f)/A_1}{A_1^2+A_2^2},
\end{eqnarray}
where $\int_a^b\rho(f)df=A_1+iA_2$.
\newline
Now we may wonder whether the value of $Re \la f\ra=\int_a^b f w_1(f)df$ would give us an 
idea  about the location of the interval $[a,b]$. From Eq.(\ref{C2}) we note that if both 
$\rho_1(f)$ and $\rho_2(f)$ do not change sign, $w_1(f)$ is a proper probability distribution, and 
its mean certainly lies within the region of its support. If, on the other hand,  both 
$\rho_1(f)$ and $\rho_2(f)$ alternate, the mean $Re \la f\ra$ is allowed to lie anywhere, and is not 
obliged to tell us anything about the actual range of values of $f$.
\newline
So here is how a confusion might arise: suppose one needs to evaluate the average of a variable known to take
values between $1$ and $2$ indirectly, i.e., without checking whether the distribution alternates, or is a proper probabilistic
one.  Obtaining a result of $-9$ may seem unusual, until it is realised that the employed distribution changes sign, and 'scrambles' 
the information about the actual range values involved. 
\newline
One remaining question is why was it necessary to employ such a tricky distribution in the first place?

  \section{Feynman's Uncertainty Principle and the 'Which way?' question}
 A chance to employ oscillatory complex valued distribution is offered by quantum mechanics, 
 and for a good reason.
 Consider a kind of double-slit experiment in which a quantum system, initially in a state $|I\ra$,  may reach a given 
 final state $|F\ra$ via two pathways, the corresponding probability amplitudes being $A(1)$ and $A(2)$.
 There are two possibilities.
 \newline
 (I) The pathways interfere, and the probability to reach $|F\ra$ is given by 
 \begin{equation} \label{U1}
P^{F\leftarrow I} = |A(1)+A(2)|^2.
\end{equation}
\newline
(II) Interference between the pathways has been completely destroyed by bringing the system in contact with
another system, or an environment. Now the probability to reach $|F\ra$ is
  \begin{equation} \label{U2}
P^{F\leftarrow I} = |A(1)|^2+|A(2)|^2.
\end{equation}
The two cases are physically different, as are the two probabilities. 
 In the second case the two pathways are ${\it real}$. 
 One can make an experiment which would confirm by multiple trials that the system travels 
 either the first or the second route with frequencies  proportional to $|A(1)|^2$ and $|A(2)|^2$, respectively.
In the first case the pathways remain ${\it virtual}$. Together they form a single 
real pathway travelled with probability $|A(1)+A(2)|^2$, and there is no way of 
saying, even statistically,  which of the two virtual paths the system has actually travelled.

The above leads to a loose formulation of the Uncertainty Principle \cite{Feyn}: several interfering pathways or states
should be considered as a single unit. Quantum interference erases detailed information about a system.
This information can only be obtained if interference is destroyed, usually at the cost of perturbing
the system's evolution, thus destroying also the very studied phenomenon, e.g., an interference pattern 
in Young's double-slit  experiment.

 \section{Feynman paths and pathways}
Let us  go about the pathways in a slightly more formal way.
By slicing the time interval into $N$ subintervals, and sending $N$ to infinity,
we can write the 
 transition amplitude for a system with a Hamiltonian $\hat{H}$ as a sum 
over paths traced by a variable $\hat{A}$,
\begin{eqnarray} \label{F1}
\la F|\exp(-i\hat{H}t/\hbar)|I\ra 
= lim_{N\rightarrow \infty}\sum_{k_{1},k_{2},...k_{N+1}}\times\\ \nonumber
\la F|a_{k_{N+1}}\ra
\la a_{k_{N+1}}|exp(-i\hat{H}t/\hbar N)|a_{k_{N}}\ra\la a_{k_{N}}|...|a_{k_{2}}\ra
\la a_{k_{2}}|exp(-i\hat{H}t/\hbar N)|a_{k_{1}}\ra\la a_{k_{1}}|I\ra
\\
\nonumber
\equiv \sum_{paths} A^{F\leftarrow I}[path]
\end{eqnarray}
where $a_k$ and $|a_k\ra$ are
the eigenvalues and eigenvectors of the variable of interest $\hat{A}$,
%
$\hat{A}|a_k\ra = a_k|a_k\ra$.
We also introduced Feynman paths - functions which take the values $a_k$ from the spectrum of $\hat{A}$
at each discrete time.  In the limit $N\to \infty$ we will denote such a path by $a(t)$. The paths are virtual pathways, each contributing a probability amplitude $A^{F\leftarrow I}[path]$ defined in Eq.(\ref{F1}).
In the chosen representation they form the most detailed complete set of histories available to the quantum system.
\newline
We may be interested not in every detail of the particle's past, but only in the value of a certain variable, 
a functional defined for a Feynman path $a(t)$ as an integral
\begin{equation} \label{F1a}
 \mathcal{F}[path]= \int_{0}^t \beta(t') a(t') dt', 
\end{equation} 
 where $\beta(t)$ is
a known function of our choice. We can define a less detailed set of virtual pathways by grouping together those paths 
for which the value of $\mathcal{F}[a]$ equals some $f$. Each pathway now contributes the amplitude
\begin{equation} \label{F2}
\Phi^{F\leftarrow I}(t|f)= \sum_{paths} \delta(f-\mathcal{F}[path])
A^{F\leftarrow I}[path],
\end{equation}
where $\delta(z)$ is  the Dirac delta. The new pathways contain the most detailed information about the variable $\mathcal{F}$, 
while information about other variables has been lost to interference in the sum (\ref{F2}).
\newline
Next we can define a coarse grained amplitude distribution for $\mathcal{F}$ by smearing $\Phi^{F\leftarrow I}(t|f)$ with a 'window'
function $G(f)$
:
\begin{equation} \label{F3}
\Psi^{F\leftarrow I}(t|f) = \int G(f-f')\Phi^{F\leftarrow I}(t|f') df'.
\end{equation}
With $G(f)$ chosen, for example,  to be a Gaussian of a width $\Delta f$ we are unable to distinguish the values $f_1$ and $f_2$ less than $\Delta f$ apart,
$|f_1-f_2|\lesssim \Delta f$, since the corresponding pathways may now interfere. 
\newline
The coarse graining does, however, have 
a physical meaning. Consider a basis $\{F\}$ containing our final state $|F\ra$, and construct a state
$|\Psi^{I}(t|f)\ra \equiv \sum_F |F\ra \Psi^{F\leftarrow I}(t|f)$ so that $ \Psi^{F\leftarrow I}(t|f)=\la F|\Psi^{I}(t|f)\ra$.
It is easy to check \cite{DS2013} that $\Psi^{I}(t|f)\ra$ satisfies a differential equation,
\begin{equation} \label{F4}
i\partial_t |\Psi^{I}(t|f)\ra = [\hat{H} - i\hbar \partial_f \beta(t)\hat
{A}|\Psi^{I}(t|f)\ra
\end{equation}
with the initial condition
\begin{equation} \label{F5}
|\Psi^{I}(t=0|f)\ra=G(f)|I\ra,
\end{equation}
This can also be seen as a Schroedinger equation describing a system interacting
with a von Neumann pointer \cite{vN} whose position is $f$. With it we have the recipe for  measuring the the quantity $\mathcal{F}[path]$:
first prepare the system in the initial state $|I\ra$  and the pointer in the state $\int G(f)|f\ra df$. 
Switch on the coupling, and at a time $t$ measure the pointer position accurately.
Interference between paths with different values of $\mathcal{F}[path]$ will be destroyed, 
since they lead do different outcomes for the pointer.

 \section{The accuracy and the back action}
Our measurement scheme has an important parameter, the width of the 
window $G(f)$, $\Delta f$, which determines the extent to which we
can ascertain the value of $\mathcal{F}[path]$, once the pointer has been 
found in $f$. This accuracy parameter also determines the perturbation 
a measurement exerts on the measured system. This, in turn, can be judged 
by how much the probability to arrive in a final state $|F\ra$ with the 
meter on differs from that with the meter off. The former is given by 
\begin{equation} \label{W1}
P^{F\leftarrow I}(t)=\int df |\Psi^{F\leftarrow I}(t|f)|^2,
\end{equation}
and, in general, is not equal  to $|\la F|\exp(-i\hat{H}t/\hbar)|I\ra|^2 $
since
\begin{equation} \label{W2}
\int G(f-f')\Phi^{F\leftarrow I}(t|f')df' \ne G(f) \la F |\exp (-i\hat{H}t/\hbar)|I\ra
= G(f)\int\Phi^{F\leftarrow I}(t|f')df',
\end{equation}
where the last equality is obtained by integrating Eq.(\ref{F2}).
\newline
The perturbation can be minimised by choosing $G(f)$ to be very broad.
By construction, the value of  $\mathcal{F}$ typically lies within a finite interval, 
say, $a \le \mathcal{F}[path] \le b$, outside of which $\Phi^{F\leftarrow I}(t|f')$ vansihes.
A very broad $G(f-f')$ can, therefore  be replaced by $G(f)$, making the l.h.s.
of Eq.(\ref{W2}) proportional to $\la F|\exp (-i\hat{H}t/\hbar)|\Psi_0\ra$.
\newline
Thus, in order to study the system with the interference between the pathways 
intact, we must make a highly inaccurate 'weak' measurement.
This can be achieved by introducing a high degree of uncertainty in 
the pointer's initial position. The following classical example may give 
us some encouragement.

 \section{Inaccurate classical measurements}
Consider a classical system which can reach a final 
state by several different routes. Let us say, a ball can roll  
from a hole $I$ to a hole $F$  down the first groove with
the probability $w_1>0$, or down the second groove, with 
the probability $w_2>0$, and so on. It is easy to imagine a (purely 
classical) pointer which moves one unit to the right
if the ball travels the first route, or two units to the right, 
if the second route is travelled, and so on. The meter is imperfect:
we can accurately  determine its final position, while 
we cannot be sure that it has been set exactly at zero.
Rather, its initial position is distributed around $0$ with a probability density 
$G(f)$ of a zero mean and a known variance. 
Let there be just two routes. Now the final meter readings are also uncertain, 
with the probability to find it in $f$ given by
\begin{eqnarray} \label{I1}
P^{F\leftarrow I}(f)=\int G(f-f')w(f')df'\\ \nonumber
w(f)\equiv w_1\delta(f-1)+w_2\delta(f-2).
\end{eqnarray}
If the meter is accurate, i.e., if $G(f)$ is  very narrowly peaked around $f=0$,
we will have just two possible readings, $f=1$, in approximately
$w_1N$ out of $N$ trials, or $f=2$, in approximately $w_2N$ out of 
all cases.
\newline
Suppose next that the meter is highly inaccurate, and the width 
of $G$, $\Delta f$ is much larger than $1$. A simple calculation 
shows \cite{C3} that the first two moments of the final distribution are given by 
 \begin{eqnarray} \label{I2}
\la f \ra = \int f w(f) df,\quad \quad \quad \quad \quad \quad \quad \quad \quad \quad \quad \quad \\ \nonumber
\la f^2 \ra=\int f ^2w(f) df +\int f^2 G(f)df/\int G(f)df.
\end{eqnarray}
We have, therefore, a very broad distribution, whose mean 
coincides with the mean of the $w(f)$. Since the second moment of $G$ is known, by performing a large 
number of trials we can extract from the data also the variance $\sigma$  of 
$w(f)$. For instance, if the two routes are travelled with equal probabilities, 
$w_1=w_2=1/2$, we have 
\begin{eqnarray} \label{I3}
\la f \ra=1.5, \q \sigma = 0.5.
\end{eqnarray}
From this we can correctly deduce that there are just two, and not three or four, routes
available to the system, and that they are travelled with roughly equal probabilities.
This simple example shows that, classically, even a highly inaccurate meter can yield
limited information about the alternatives available to a stochastic system.
It is just a matter of performing a large number of trials required to gather the necessary 
statistics. Next we will see whether this remains true in the quantum case.

\section {Inaccurate, or 'weak', quantum measurements,}
In the quantum case, employing an inaccurate meter has a practical advantage
- we minimise the back action of the meter on the measured system, 
and may hope to learn something without destroying the interference.
As discussed in Sect. VI, we can make a measurement non-invasive
by giving the initial meter's position a large quantum uncertainty
(that is to say, we choose a pure meter state broad in the coordinate space).
We prepare the system and the pointer in a product state 
(\ref{F5}), turn on the interaction, check the system's final state, and sample 
the meter's reading provided this final state is $|F\ra$.
From (\ref{F4}) the moments of the  distribution of the meter's readings
are given by
\begin{equation} \label{WW1}
\la f^n\ra \equiv
 \int f^n| \Psi^{F\leftarrow I}(t|f)|^2 df/\int |\Psi^{F\leftarrow I}(t|f)|^2 df.
\end{equation}
 As the width of the initial  meter's state  $\Delta f $ tends to infinity, assuming 
$ImG(f)=0$
 we have \cite{C3} 
%
\begin{equation} \label{WW2}
\la f \ra  = Re \bar{f} + O(1/\Delta f),
\end{equation}
and
\begin{equation} \label{WW3}
\la f^2 \ra = \frac{\int f^2G(f)^2 df}{\int G(f)^2 df}
+C(Re \bar{f^2}-|\bar{f}|^2)+|\bar{f}|^2+ O(1/\Delta f).
\end{equation}
where $C$ is a factor of order of unity, which depends only on the shape of $G(f)$ \cite{C3}.
%
%
and we have introduced the notation $\bar{f^n}$ for the $n$-th moment
of the complex valued
amplitude distribution $ \Phi(f)$ defined in Eq.(\ref{F2}),
\begin{equation} \label{WW5}
\bar{f^n}\equiv
 \int f^n \Phi^{F\leftarrow I}(t|f) df/\int \Phi(t|f) df.
\end{equation}
It is at this point that 'improper' averages (\ref{WW5}) evaluated with oscillatory distributions enter
our calculation, originally set to evaluate 'proper' probabilistic averages (\ref{WW2}).
Expressions similar to Eq.(\ref{WW2}) have been
obtained earlier in  \cite{W1,W4} for a weak von Neumann measurement
and in \cite{SB} for the quantum traversal time. They are the quantum 
analogues of the classical Eqs.(\ref{I2}).
\newline
We see that the quantum case turned out to be different in one important aspect.
Where the inaccurate classical calculation of the previous Section yielded the mean of the {\it probability distribution}, its quantum counterpart gives us the mean evaluated with the {\it probability amplitude}
$ \Phi^{F\leftarrow I}(t|f)$.
There is no {\it apriori} reason to expect that either its real or imaginary part does not change
sign. As discussed in Sects. II and  III, such averages are not obliged to tell us anything about 
the actual range of a random variable. Thus, our attempt to answer the 'Which way?' ('Which $f$?') 
question is likely to fail, as we are not able to extract the information about the alternatives available to 
a quantum system. But we have been warned: the Uncertainty Principle suggests that, 
for as long as the pathways remain interfering alternatives, the question we ask has no meaning.

\section {A double slit experiment}
To give our approach  a concrete example, we return to the double slit experiment. Consider a two-level system, 
e.g., a spin-1/2 precessing in a magnetic field. The Hamiltonian is given by
\begin{equation} \label{5.1}
\hat{H}= \hbar \omega_L \sigma_x
\end{equation}
where $\omega_L $ is the Larmor frequency, and $\sigma_x$ is the Pauli matrix.
We assume that the spin is pre-selected in a state polarised along the $z$-axis at $t=0$, 
and then post-selected in the same state at  $t=T$. We also wish to know 
the state of the spin half-way through the transition, at $t=T/2$.
We follow the steps outlines in Sect. V.
At any given time, and in the given representation, the spin can point
up or down the $z$-axis.
We label these two sates $|1\ra$ and $|2\ra$, respectively.
Feynman paths are, therefore, irregular curves shown in Fig. 1.
\begin{figure}[ht]
\epsfig{file=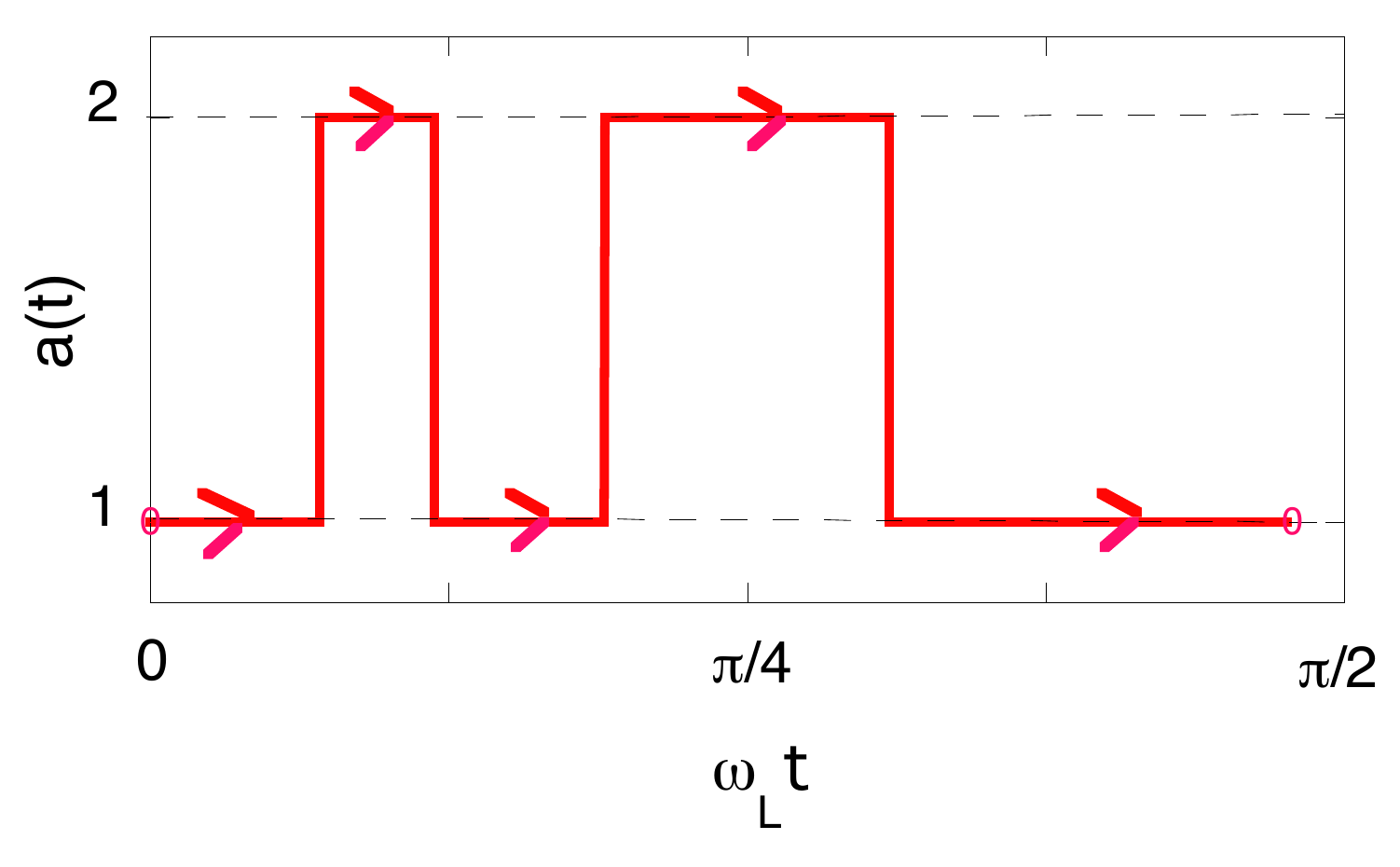, height=8cm, angle=0}
\vspace{0pt}
\caption{
\newline
Schematic diagram showing a Feynman path $a(t)$ for a spin-1/2 precessing in a magnetic field.
The path connects the state $|1\ra$ at $t=0$ with the same state at $t=2/\omega_L$.
Between these times the path jumps between $1$ and $2$, passing through $|2\ra$ at $t=1/\omega_L$.}
\end{figure}
The functional $\mathcal{F}(path)$ is given by Eq.(\ref{F1a}) with 
$\beta(t')=\delta(t'-T/2)$,
\begin{equation} \label{D1}
\mathcal{F}(path)=\int_0^t \delta(t'-T/2) a(t')dt'=a(T/2).
\end{equation}
Thus, we combined the Feynman paths ending in the state $|1\ra$ at $t=T$ into two virtual pathways,
one containing the paths passing at $t=T/2$ through the 
state $|1\ra$, and the other - the paths passing through the state 
 $|2\ra$. The corresponding probability amplitudes are those 
 for evolving the spin freely from its initial state to $|1\ra$ or $|2\ra$
 at $t=T/2$, and then to the final state $|1\ra$ at $t=T$, 
\begin{eqnarray} \label{D2}
A(1)=\cos^2(\omega_L T/2)\q
\\
\nonumber
A(2)=-\sin^2(\omega_L T/2).
\end{eqnarray}  
We will need a meter.
The interaction $-i\partial_f\delta(t-T/2)\hat{A}$ corresponds to
a von Neumann measurement \cite{vN} of the operator $\hat{A}=1\times |1\ra\la1|+2\times |2\ra\la2|$
performed at $t=T/2$. The accuracy of the measurement depends on the width $\Delta f$ of the initial 
meter's state, which we will choose to be a Gaussian,
\begin{equation} \label{D3}
G(f) = (2/\pi\Delta f^2 )^{1/4}\exp(-f^2/\Delta f^2), \q \int|G(f)|^2 df =1.
\end{equation}
It is easy to check that the average meter reading $\la f\ra$ in Eqs.(\ref{I2}) is given by
 \begin{eqnarray} \label{D3a}
\la f\ra=\frac{A(1)^2+2A(2)^2+3A(1)A(2)\exp(-0.5/\Delta f^2)}{A(1)^2+A(2)^{2}+2A(1)A(2)\exp(-0.5/\Delta f^2)},
\end{eqnarray} 
its dependence on $\Delta f$ shown in Fig.2.
\begin{figure}[ht]
\epsfig{file=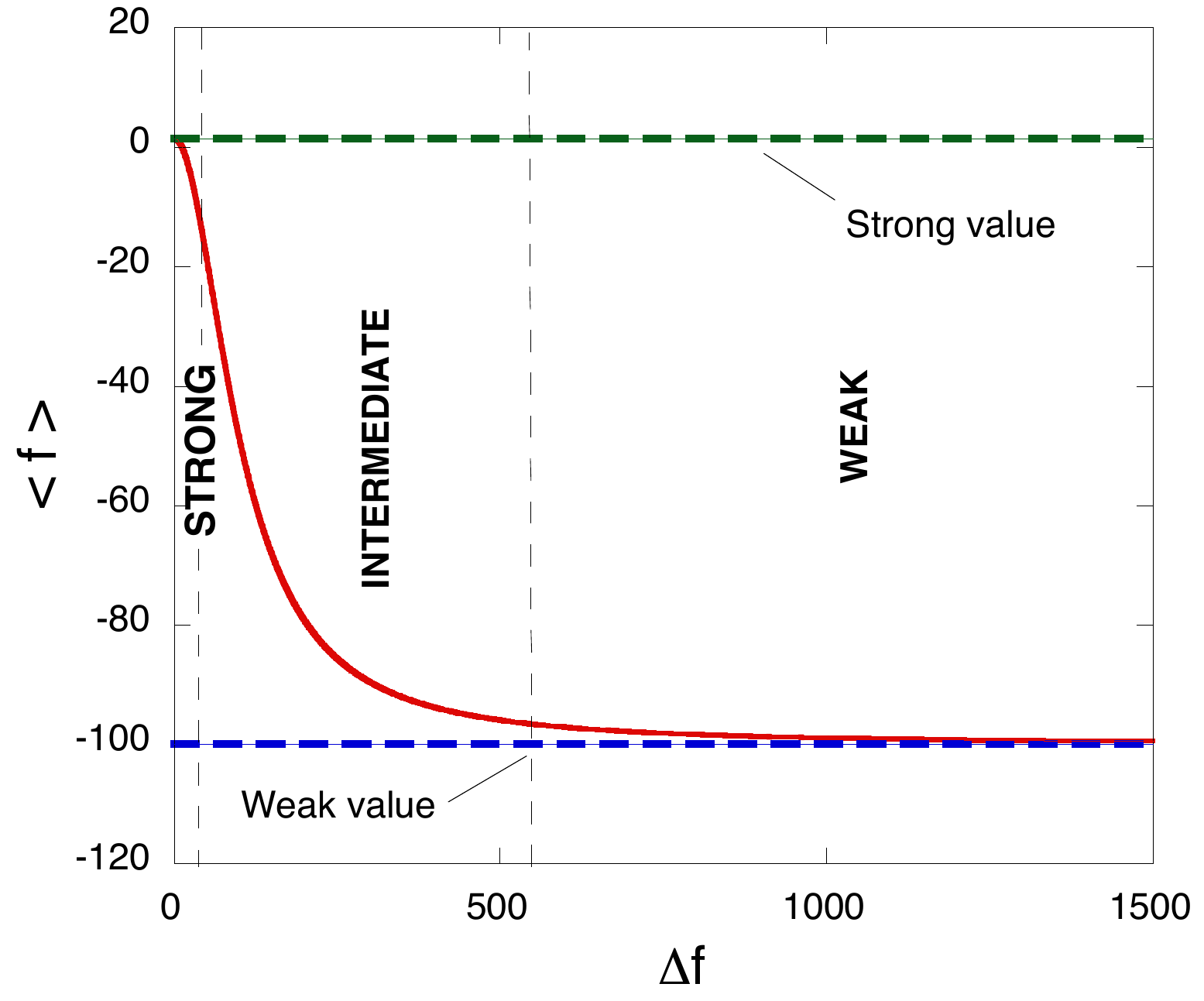, height=11cm, angle=0}
\vspace{0pt}
\caption{The mean meter reading as the function of the accuracy of the measurement in the double-slit case of the Sect. IX.
In the strong measurement regime the meter destroys coherence between the pathways
passing through different slits, but also destroys the interference pattern.
In the weak regime the interference is intact, but the measured mean slit number is 
$-100$. In the intermediate regime the mean slit number changes smoothly from $1.5$ to $-100$.
}
\end{figure}
\newline
This is, of course,  an oversimplified version of the Young's double slit experiment:
the states at $t=T/2$ play the role of the two slits, and the states at $t=T$ - the role of the 
positions on the screen where an 'interference pattern' is observed.
\newline
Consider first a 'strong' measurement of the slit number. Choose the final time such that 
finding the freely precessing spin in the state $|1\ra$ is unlikely (our 'interference pattern' has there a minimum, or a 'dark fringe'), say
$T= \arccos(1/203)/\omega_L\approx 1.5659/\omega_L$. 
Sending $\Delta f \to 0$, for the probability distribution of the meter's readings
we have [cf. Eq.(\ref{W1})]
\begin{eqnarray} \label{D4}
P^{1\leftarrow 1}(T|f)=\cos^4(\omega_L T/2)\delta(f-1) + \sin^4(\omega_L T/2)\delta(f-2) \\ \nonumber
\approx 0.252\delta(f-1) + 0.248\delta(f-2).
\end{eqnarray}
We observe that the two pathways are travelled with almost equal probability, and Eq.(\ref{D3a})
gives us the mean slit number 
 $$ \la f \ra_{strong} \approx 1.5.$$
However, this is not the original spin precession we set out to study. The interference 
pattern has been destroyed and the probability to arrive at the final position $|1\ra$, 
which without a meter was
\begin{equation} \label{D5}
|A(1)+A(2)|^2 \approx  0.000024, 
\end{equation}
is now close to $0.5$.
This is a textbook example which illustrates 
the Uncertainty  Principle: converting virtual paths into real ones comes at the cost 
of loosing the interference pattern.

Not satisfied, we try to minimise the perturbation in the hope to 
learn something about the route chosen by the system with the interference intact. 
 We send $\Delta f$ to infinity, and after many trials obtain the answer: the mean 
 number of the slit used is 
\begin{equation} \label{D6}
\la f \ra_{weak} = Re \frac{1\times A(1)+2\times A(2)}{A(1)+A(2)}= -100.
\end{equation}
Which brings us back to our original question, to rephrase the title 
of Ref.\cite{W1}, "How the result of measuring the number 
of the slit in a double slit experiment can turn out to be $-100$?" 
\section{Conclusions and discussion}
We have tried to evaluate the mean number of the slit a particle 
goes through in a double slit experiment, and came up with the 
number $-100$. The mathematics is straightforward, and we need 
to understand the meaning of this result before employing the 'weak measurements' elsewhere. There are just 
two slits, numbered $1$ and $2$,  so the result looks a bit strange.
Has our measurement gone wrong, or is the quantum world so 
strange that there are slits we are not aware of?
We opt for the first choice. 

Wrong measurements are common
in classical physics. They can be made and  repeated, but only have
meaning within the narrow context of the wrong experiment itself.
A broken speedometer will read $50$ m.p.h each time the car 
goes at $100$ m.p.h, and might convince the driver, but not 
the traffic policemen who stops him for speeding.
The slit number $-100$ may come up in a weak measurement, but 
cannot be used for any other purpose, such as convincing  a potential
user that the screen he is about to buy has more than two holes 
drilled in it.
\newline
There is, however, one important distinction. 
Classically, one can always find the right answer and correct, 
or re-calibrate the errant speedometer. Quantally, it is not so.
According to the Uncertainty Principle, there is no correct answer to 
the question asked. 
The nearest classical analogy might be this. Suppose a (purely classical)
charge can be transferred across one of the two lead wires, and an observer 
can measure, which one has been chosen. Then the wires are heated up and melted into 
one. Which of the two wires has the charge gone through now?
This is what interference does, it 'melts'  the pathways through the two slits
into a single one, thus depriving the 'Which way?' question of its meaning.
\newline
Having started to use analogies it is difficult to stop. Here is the last one: one  
asks a manager a question the said manager is unable or unwilling 
to answer properly. Yet an answer he/she must give. The answer (or no-answer) given will have 
little to do with what one wants to know. It will be repeated should the 
question be asked again. It shouldn't, however, be used to draw further 
conclusions about the matter of interest.
\newline
The 'weak measurements' rely on an interesting interference effect which has applications
beyond measurement theory \cite{HART}, \cite{LAM}. They can be made, and
have been made in practice \cite{W2}. They have useful applications in 
interferometry \cite{W7,W8}. However, their results should not be over-interpreted.
Bizarre weak values indicate the failure of a measurement procedure under
the conditions where, according to the Uncertainty Principle, it must fail.
Seen like this, the ' weak measurements' loose much of their original appeal, 
and the calculation of 'weak values' reduces to a simple exercise in first order 
perturbation theory. 
\newline
Finally, throughout the paper we appealed to the Uncertainty Principle, seen as one of the basic axioms of quantum theory. It is possible that the Principle itself will be explained in simpler terms within a yet unknown general theory. However, we argue, that the weak measurements have not yet given such an explanation, nor provided any deeper insight into physical reality.
\acknowledgements
I acknowledge support of the Basque Government (Grant No. IT-472-10), and the Ministry of Science and Innovation of Spain (Grant No. FIS2009-12773-C02-01). I am also grateful to Dr. G. Gribakin for bringing the lines used in the epigraph to my attention. 


\begin{thebibliography}{References}
\bibitem{W1}Aharonov Y, Albert DZ, Vaidman L. How the result of a
measurement of a component of the spin of a spin-$\frac{1}{2}$ particle
can turn out to be 100. Physical Review Letters 1988; 60 (14): 1351--1354.
\url{http://dx.doi.org/10.1103/PhysRevLett.60.1351}

\bibitem{W2}Duck IM, Stevenson PM, Sudarshan ECG. The sense in which
a ``weak measurement'' of a spin-$\frac{1}{2}$ particle's spin component
yields a value 100. Physical Review D 1989; 40 (6): 2112--2117. \url{http://dx.doi.org/10.1103/PhysRevD.40.2112} 

\bibitem{W3}Ritchie NWM, Story JG, Hulet RG. Realization of a measurement
of a ``weak value''. Physical Review Letters 1991; 66 (9): 1107--1110.
\url{http://dx.doi.org/10.1103/PhysRevLett.66.1107}

\bibitem{W4}Aharonov Y, Vaidman L. The two-state vector formalism
of quantum mechanics. In: Time in Quantum Mechanics, Muga G, Mayato
RS, Egusquiza I (editors), Springer, 2002, pp.369--412. \url{http://arxiv.org/abs/quant-ph/0105101}

\bibitem{W5}Aharonov Y, Botero A, Popescu S, Reznik B, Tollaksen
J. Revisiting Hardy\textquoteright{}s paradox: counterfactual statements,
real measurements, entanglement and weak values. Physics Letters A
2002; 301: 130--138. \url{http://arxiv.org/abs/quant-ph/0104062}

\bibitem{W6}Jozsa R. Complex weak values in quantum measurement.
Physical Review A 2007; 76 (4): 044103. \url{http://arxiv.org/abs/0706.4207}

\bibitem{W7}Dixon PB, Starling DJ, Jordan AN, Howell JC. Ultrasensitive
beam deflection measurement via interferometric weak value amplification.
Physical Review Letters 2009; 102 (17): 173601. \url{http://arxiv.org/abs/0906.4828}

\bibitem{W8}Popescu S. Weak measurements just got stronger. Physics
2009; 2: 32. \url{http://dx.doi.org/10.1103/Physics.2.32}

\bibitem{W9}Dressel J, Jordan AN. Sufficient conditions for uniqueness
of the weak value. Journal of Physics A: Mathematical and Theoretical
2012; 45 (1): 015304. \url{http://arxiv.org/abs/1106.1871}

\bibitem{W10}Rozema LA, Darabi A, Mahler DH, Hayat A, Soudagar Y,
Steinberg AM. Violation of Heisenberg\textquoteright{}s measurement-disturbance
relationship by weak measurements. Physical Review Letters 2012; 109
(10): 100404. \url{http://dx.doi.org/10.1103/PhysRevLett.109.100404}

\bibitem{BBC}Palmer J. Heisenberg uncertainty principle stressed
in new test. BBC News: Science \& Environment. Publication date: 7
September 2012; \url{http://www.bbc.co.uk/news/science-environment-19489385}

\bibitem{C1}Leggett AJ. Comment on ``How the result of a measurement
of a component of the spin of a spin-$\frac{1}{2}$ particle can turn
out to be 100''. Physical Review Letters 1989; 62 (19): 2325--2325.
\url{http://dx.doi.org/10.1103/PhysRevLett.62.2325}

\bibitem{C2}Peres A. Quantum measurements with postselection. Physical
Review Letters 1989; 62 (19): 2326--2326. \url{http://dx.doi.org/10.1103/PhysRevLett.62.2326}

\bibitem{C3}Sokolovski D. Weak values, ``negative probability,''
and the uncertainty principle. Physical Review A 2007; 76 (4): 042125.
\url{http://arxiv.org/abs/0905.3810} 

\bibitem{C4}Sokolovski D, Puerto Gim\'enez I, Sala Mayato R. Feynman-path
analysis of Hardy's paradox: Measurements and the uncertainty principle.
Physics Letters A 2008; 372 (21): 3784--3791. \url{http://arxiv.org/abs/0903.4795}

\bibitem{C4b}Sokolovski D, Puerto Gim\'enez I, Sala Mayato R. Path
integrals, the ABL rule and the three-box paradox. Physics Letters
A 2008; 372 (44): 6578--6583. \url{http://arxiv.org/abs/0903.4600}

\bibitem{C5}Parrott S. Quantum weak values are not unique. What do
they actually measure? 2009; \url{http://arxiv.org/abs/0909.0295}

\bibitem{Feyn}Feynman RP, Leighton RB, Sands M. The Feynman Lectures
on Physics, Volume 3. Reading, Massachusetts: Addison-Wesley, 1965.

\bibitem{Bohm}Bohm D. Quantum Theory. New York: Dover Publications,
1989., p. 600.

\bibitem{DS2013}Sokolovski D. Path integral approach to space-time
probabilities: A theory without pitfalls but with strict rules. Physical
Review D 2013; 87 (7): 076001. \url{http://arxiv.org/abs/1301.1244}

\bibitem{vN}von Neumann J. Mathematical Foundations of Quantum Mechanics.
Investigations In Physics, Beyer RT (translator), Princeton: Princeton
University Press, 1955, pp.183--217.

\bibitem{SB}Sokolovski D, Baskin LM. Traversal time in quantum scattering.
Physical Review A 1987; 36 (10): 4604--4611. \url{http://dx.doi.org/10.1103/PhysRevA.36.4604}

\bibitem{HART}Sokolovski D, Akhmatskaya E. Hartman effect and weak
measurements that are not really weak. Physical Review A 2011; 84
(2): 022104. \url{http://arxiv.org/abs/1103.5620}

\bibitem{LAM}
Monks PDD, Xiahou C, Connor JNL. Local angular momentum-local impact parameter analysis: Derivation and properties of the fundamental identity, with applications to the F + H2, H + D2, and Cl + HCl chemical reactions. Journal of Chemical Physics 2006; 125 (13): 133504-133513.  \url{http://dx.doi.org/10.1063/1.2210480}
 \end{thebibliography}
\end{document}